\documentclass[aps,prl,twocolumn,superscriptaddress]{revtex4}

\usepackage{graphicx}
\usepackage{bm}

\begin{document}

\preprint{}

\title{Antiferromagnetic Spin Fluctuations and Unconventional Nodeless Superconductivity in an Iron-based New Superconductor (Ca$_4$Al$_2$O$_{6-y}$)(Fe$_2$As$_2$):$^{75}$As-NQR Study}

\author{H. Kinouchi}
\email[]{e-mail  address: kinouchi@nmr.mp.es.osaka-u.ac.jp}
\affiliation{Graduate School of Engineering Science, Osaka University, Osaka 560-8531, Japan}

\author{H. Mukuda}
\email[]{e-mail  address: mukuda@mp.es.osaka-u.ac.jp}
\author{M. Yashima}
\affiliation{Graduate School of Engineering Science, Osaka University, Osaka 560-8531, Japan}
\affiliation{JST, TRIP (Transformative Research-Project on Iron Pnictides), Chiyoda, Tokyo 102-0075, Japan}

\author{Y. Kitaoka}
\affiliation{Graduate School of Engineering Science, Osaka University, Osaka 560-8531, Japan}

\author{P. M. Shirage}
\author{H. Eisaki}
\author{A. Iyo}
\affiliation{National Institute of Advanced Industrial Science and Technology (AIST), Umezono, Tsukuba 305-8568, Japan}
\affiliation{JST, TRIP (Transformative Research-Project on Iron Pnictides), Chiyoda, Tokyo 102-0075, Japan}

\date{\today}

\begin{abstract}
We report $^{75}$As-nuclear quadrupole resonance (NQR) studies on (Ca$_4$Al$_2$O$_{6-y}$)(Fe$_2$As$_2$) with $T_{\rm c}=27$~K, which unravel unique normal-state properties and point to unconventional nodeless superconductivity (SC). Measurement of nuclear-spin-relaxation rate $1/T_1$ has revealed a significant development of two dimensional (2D) antiferromagnetic (AFM) spin fluctuations down to $T_{\rm c}$, in association with the fact that FeAs layers with the smallest As-Fe-As bond angle are well separated by thick perovskite-type blocking layer. Below $T_{\rm c}$, the temperature dependence of $1/T_1$ without any trace of the coherence peak is well accounted for by an $s_{\pm }$-wave multiple gaps model. From the fact that $T_{\rm c}=27$~K in this compound is comparable to $T_c$=28 K in the optimally-doped LaFeAsO$_{1-y}$ in which AFM spin fluctuations are not dominant, we remark that AFM spin fluctuations are not a unique factor for enhancing $T_c$ among existing Fe-based superconductors, but a condition for optimizing SC should be addressed from the lattice structure point of view.
\end{abstract}


\maketitle


Iron-based high-$T_{\rm c}$ superconductors\cite{Kamihara} comprise a two-dimensional layered structure of iron (Fe)-pnictgen ({\itshape Pn}) planes, which are separated by blocking layers, such as {\itshape Ln}O ({\itshape Ln}=rare earth), alkaline earth, and alkaline metal elements. 
Relatively high SC transition has been reported in Fe-pnictides with a thick perovskite-type blocking layer, in which the interlayer distance between Fe{\itshape Pn} layers is more than $\sim 13$\,\AA\ \cite{Zhu,Ogino1,Ogino2,Shirage}. For example, $T_{\rm c}$ is $\sim 47$\,K for (Ca$_4$(Mg$_{0.25}$Ti$_{0.75}$)$_3$O$_y$)(Fe$_2$As$_2$)\cite{Ogino2}, and $\sim 37$\,K for (Sr$_4$V$_2$O$_6$)(Fe$_2$As$_2$)\cite{Zhu}, which rises up to $46$\,K by the application of pressure\cite{Kotegawa}.  
Moreover, in these series of Fe-based compounds, neither structural transition nor magnetic order has been reported so far, differentiating them from other Fe-based superconductors that emerge in close proximity to antiferromagnetic (AFM) order\cite{Kamihara}. In fact, the maximum of SC transition temperature $T_{\rm c}$ seems to take place around a quantum-critical point (QCP) of AFM order for Ba(Fe$_{1-x}$Co$_{x}$)$_2$As$_2$\cite{Ning} and BaFe$_2$(As$_{1-x}$P$_{x}$)$_2$\cite{Kasahara,Nakai}. On the one hand, the $T_c$ of Fe-pnictides is intimately related with local structural parameters such as a $Pn$-Fe-$Pn$ bond angle of Fe$Pn_4$ tetrahedron ({\it Lee's plot})\cite{C.H.Lee} and/or a height of pnictgen from Fe$Pn$-plane\cite{Mizuguchi}. In this context, systematic investigations on Fe-based superconductors with a thick blocking layer are required in order to get insight into some correlation between $T_{\rm c}$ and structural parameters and/or AFM spin fluctuations. 

In this Letter, we report $^{75}$As-nuclear quadrupole resonance (NQR) studies on (Ca$_4$Al$_2$O$_{6-y}$)(Fe$_2$As$_2$) with $T_{\rm c}=27$~K that unravel the development of significant AFM spin fluctuations and point to unconventional nodeless superconductivity (SC). 


A polycrystalline sample of (Ca$_4$Al$_2$O$_{6-y}$)(Fe$_2$As$_2$) with a nominal content of $y\sim0.215$ (denoted as Al-42622 hereafter) was synthesized by solid-state reaction method using high-pressure synthesis technique as described elsewhere\cite{Shirage}. Powder x-ray diffraction measurement indicates that this sample is almost composed of a single phase with lattice parameters, $a$=3.71~\AA\ and $c$=15.40~\AA. This compound is characterized by a small $a$-axis length, a narrow {\itshape Pn}-Fe-{\itshape Pn} bond angle ($\alpha $ $\sim $ $102.1^\circ$), and a high {\itshape Pn} distance from Fe$Pn$ plane with $h_{Pn}$ $\sim $ $1.50$\,\AA\cite{Shirage}. $T_{\rm c}=27$~K was determined from the onset of SC diamagnetism in susceptibility measurement. $^{75}$As-NQR measurements have been performed on a coarse powder sample at zero external field. 


\begin{figure}[h]
\centering
\includegraphics[width=6.5cm]{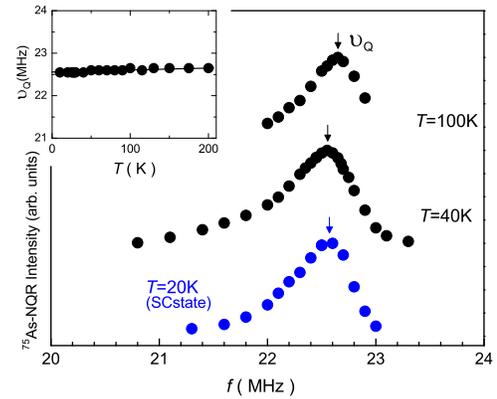}
\caption[]{(Color online) $^{75}$As-NQR spectra of Al-42622. 
The inset shows $T$ dependence of $^{75}\nu_{\rm Q}$ of Al-42622, indicating that neither structural phase transition nor no magnetic order takes place in Al-42622.
}
\label{NQRspectra}
\end{figure}

Figure~\ref{NQRspectra}(a) shows $^{75}$As-NQR spectra of Al-42622. $^{75}$As-NQR frequency ($^{75}\nu _{\rm Q}$) is  $\sim 22.6$~MHz, which is the largest among Fe-based superconductors 
so far. In $Ln$FeAsO$_{\delta}$ ($Ln$:Rare earth, denoted as $Ln$-1111 hereafter), note that $^{75}\nu _{\rm Q}$ becomes large when an $a$-axis length decreases\cite{MukudaNQR,MukudaPhysC}. Since $^{75}\nu _{\rm Q}$ is proportional to an electric field gradient at $^{75}$As nuclear site yielded by local distributions of on-site electron density and lattice ions around an $^{75}$As nucleus. In this context, the fact that $^{75}\nu_{\rm Q}$ in Al-42622 is the largest among other Fe-based compounds may be because its $a$-axis length is the shortest. The $^{75}$As-NQR spectrum is almost temperature ($T$) independent in a range of 10 K and 200 K, as shown in the inset of Fig.~\ref{NQRspectra}, demonstrating  that neither structural phase transition nor magnetic order takes place in Al-42622. An asymmetric shape of the $^{75}$As-NQR spectra in Al-42622 is probably caused  by some distribution of oxygen deficiency $y$.

\begin{figure}[h]
\centering
\includegraphics[width=6.5cm]{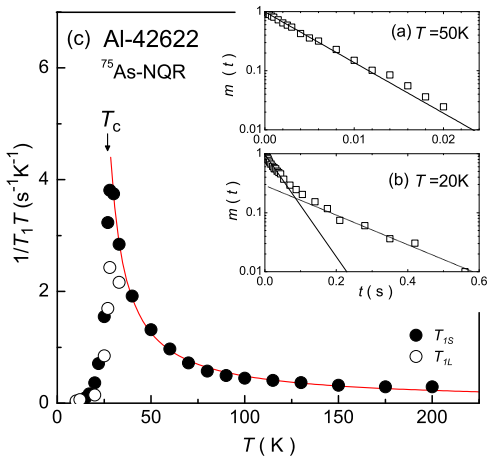}
\caption[]{(Color online) Recovery curves of $^{75}$As nuclear magnetization $m(t)$ at (a) 50 K and (b) 20 K. (c) $T$ dependence of $1/T_1T$ for Al-42622. The solid curve is a simulation fitted to a relation $1/T_{1}T \sim a/(T+\theta)+b$ 
with parameters $a$=37, $\theta $=$-20$\,K, and $b$=0.023. 
}
\label{1/T1TvsT}
\end{figure}

\begin{figure}[h]
\centering
\includegraphics[width=6.5cm]{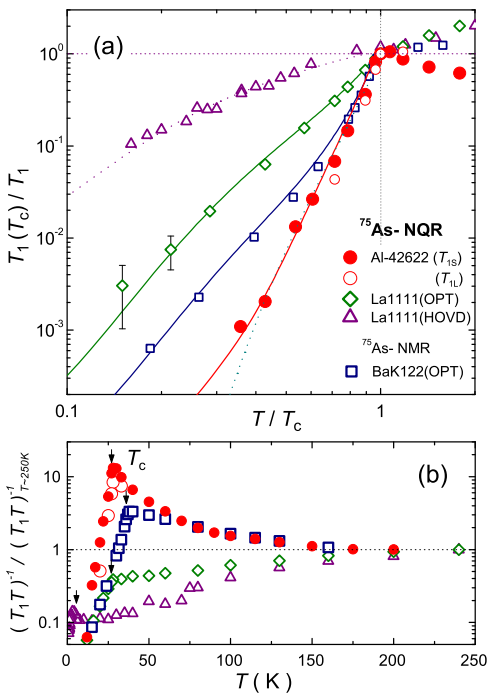}
\caption[]{(Color online) (a)  Plots of $^{75}$As-NQR $T_1(T_c)/T_1$ normalized at $T_c$ against $T/T_c$ for Al-42622, along with the results of BaK122(OPT) with $T_c$=38 K\cite{Yashima}, La1111(OPT) with $T_c$=28 K \cite{MukudaNQR}, and La1111(HOVD) with $T_c$=5 K\cite{Nitta}. 
Note that $T$ dependences of  $T_{1S}$ and $T_{1L}$ normalized at $T_{\rm c}$ for Al-42622 are  almost the same below $T_{\rm c}$. The solid curves are simulations in terms of  the $s_\pm$-wave model with multiple SC gaps (see text). 
(b) $T$ dependence of $^{75}$As-(1/$T_1T$)s normalized at $T$=250 K.
}
\label{SCvsNM}
\end{figure}

$^{75}$As-NQR $1/T_1$ is obtained by fitting a recovery curve of $^{75}$As nuclear magnetization to a single exponential function $m(t)\equiv (M_0-M(t))/M_0=\exp \left(-3t/T_1\right)$ for $I$=3/2. Here $M_0$ and $M(t)$ are the 
respective nuclear magnetizations for a thermal equilibrium condition and at time $t$ after a saturation pulse. In Al-42622, $m(t)$ was reproduced by a single component of $1/T_1$ above 40 K, but not below $\sim 40$\,K, as shown in Figs.~\ref{1/T1TvsT}(a) and \ref{1/T1TvsT}(b), respectively. 
Since the short component $1/T_{1S}$ and the long one  $1/T_{1L}$ below $\sim 40$~K exhibit almost the same $T$ dependence when normalized at $T_c$ (see Fig. \ref{SCvsNM}(a)), we focus on the $T$ dependence of $1/T_{1S}$ which is a dominant component below 40 K. 

Figure~\ref{1/T1TvsT}(c) shows the $T$ dependence of $^{75}$As-NQR $1/T_1T$ for Al-42622.
The $1/T_1T$ in the normal state increases significantly upon cooling down to $T_{\rm c}$. 
The AFM spin fluctuations in Fe-based superconductors are enhanced by the nesting of hole and electron Fermi surfaces (FSs). 
In general, $1/T_1T$ is described as $1/T_1T\propto \sum_{\bm q} |A_{\bm q}|^2 \chi''({\bm q},\omega_0)/\omega_0$, where $A_{\bm q}$ is a wave-vector (${\bm q}$)-dependent hyperfine-coupling constant, $\chi({\bm q},\omega)$ a dynamical spin susceptibility, and $\omega_0$ an NQR frequency.  When a system is close to an AFM QCP, two-dimensional (2D) AFM spin-fluctuation model predicts a relation of $1/T_1T\propto \chi_{\rm Q}(T) \propto 1/(T + \theta)$ \cite{Moriya}. Here, staggered susceptibility $\chi_{\rm Q}(T)$ with an AFM propagation vector ${\bm q}$=${\bm Q}$ follows a Curie-Weiss law. Since $1/T_1T$ diverges towards $T \rightarrow 0$ when $\theta=0$, $\theta$ is a measure of how close a system is to an AFM QCP.  Actually, as shown by the solid line in Fig.~\ref{1/T1TvsT}(c), the $1/T_{1S}T$ in Al-42622 can be fitted by assuming $1/T_{1}T \sim a/(T+\theta)+b$ with parameters $a$=37, $\theta $=$-20$\,K, and $b$=0.023. 
It is unexpected that $\theta$ is negative, meaning that the staggered susceptibility would diverge toward 20K, and hence an AFM order would be anticipated below $\sim 20$~K. 
As a matter of fact, in the case of Ba(Fe$_{1-x}$Co$_{x}$)$_2$As$_2$ and BaFe$_2$(As$_{1-x}$P$_{x}$)$_2$, the AFM order sets in when $\theta$ becomes negative\cite{Ning,Nakai}. 
However, SC occurs below $T_{\rm c}$=27 K in Al-42622, instead of an AFM order. 
This is because a thick blocking layer between FeAs layers makes an interlayer magnetic coupling weak, suppressing an onset of AFM order. 
Besides, the structure consisting of perovskite blocks bonded by strong covalent bonding prevents  a structural phase transition into an orthorhombic phase. 
These might be the main reasons why an AFM order of Fe$Pn$ layers is absent in the Fe-pnictides family with the thick blocking layers.  

The band calculation for Al-42622 reported by Miyake {\it et al.} revealed that a hole FS around $\Gamma^{\prime}$ ($\pi $,$\pi$) in the unfolded FS regime appears explicitly as a result of the small $\alpha\sim~102^{\circ}$, whereas one of two-hole FSs at $\Gamma$(0,0) is missing\cite{Miyake}. 
Eventually, it is concluded that the well nested FS topology between hole FSs at $\Gamma$ and $\Gamma^{\prime}$, and electron FSs at $M$((0,$\pi$) and ($\pi$,0)) enhances a Stoner factor of antiferromagnetism in Al-42622 \cite{Usui}. This event leads to the development of AFM spin fluctuations and hence is consistent with the experiment presented here. 


Next, we address SC characteristics  emerging under the background of AFM spin fluctuations.  
Figure~\ref{SCvsNM}(a) shows a plot of $T_1(T_c)/T_{1}$ normalized at $T_{\rm c}$ against $T/T_c$, exhibiting a steep decrease upon cooling without the coherence peak just below $T_c$. The $T$ dependence of $1/T_{1}$ seems to follow a $\sim T^7$ dependence down to $\sim $0.3$T_{\rm c}$, which is quite unique as compared with the $T^3$ in optimally-doped La-1111(OPT) with $T_{\rm c}$=28 K \cite{MukudaNQR,Nakai2} and the $T^5$ in optimally-doped Ba$_{0.6}$K$_{0.4}$Fe$_2$As$_2$(BaK122(OPT)) with $T_{\rm c}$=38 K\cite{Yashima}. 
Notably, Fig.~\ref{SCvsNM}(b) shows the $T$ dependence of $1/T_{1}T$ normalized at $T$=250 K in these compounds.
We remark that as AFM spin fluctuations are more significantly enhanced, a power-law reduction in $1/T_1$ below $T_c$ becomes steeper from $\sim T^3$ to $\sim T^7$. 

\begin{figure}[h]
\centering
\includegraphics[width=7cm]{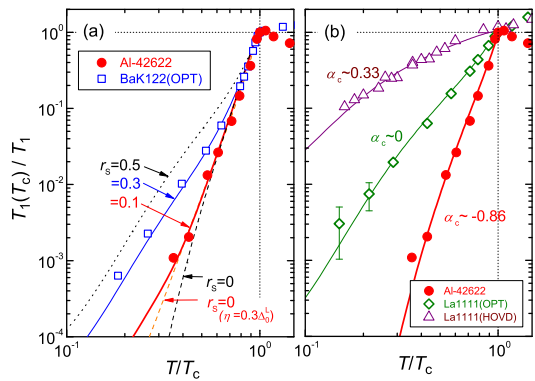}
\caption[]{(Color online) (a) Plots of $^{75}$As-NQR $T_1(T_{\rm c})/T_1$ normalized at $T_{\rm c}$ against $T/T_{\rm c}$ for Al-42622 and BaK122(OPT) with $T_{\rm c}$=38 K\cite{Yashima}. The curves are simulations in terms of the $s_\pm $-wave model with two isotropic gaps with various values of $r_{S}\equiv N_s^S/(N_s^L+N_s^S)$. Here $N_L$ and $N_S$ represent the respective DOSs with large and small SC gaps. The experimental result for Al-42622 was reproduced with $r_{S}\sim $0.1, which is smaller than $r_{S}\sim $0.3 for BaK122(OPT) \cite{Yashima}. (b) Similar plots for Al-42622, La1111(OPT) with $T_{\rm c}$=28 K\cite{MukudaNQR} and La1111(HOVD) with $T_{\rm c}$=5 K\cite{Nitta}. The experiment for Al-42622 can be also reproduced by assuming {\it a negative value} of $\alpha_{\rm c}\sim -$0.86, which contrasts with $\alpha_{\rm c}\sim 0.33$ in La1111(HOVD) and $\alpha_{\rm c}\sim 0$ in La1111(OPT).}
\label{simulation}
\end{figure}

In previous studies\cite{Yashima,Yamashita}, the non-universal $T$-dependence in $1/T_1$ was consistently accounted for by a multigap nodeless $s_{\pm }$-wave pairing model\cite{NMRtheory1,NMRtheory1a,NMRtheory1b,NMRtheory1c,NMRtheory2}. 
In the $s_\pm$-wave model with two isotropic gaps, an initial decrease of $1/T_1$ without the coherence peak just below $T_c$ is due to the opening of a large SC gap with $2\Delta _0^L/k_BT_{\rm c}$\cite{Yamashita}. 
As seen in Fig.~\ref{simulation}(a), the initial decrease in $1/T_1$ just below $T_{\rm c}$ in Al-42622 is similar to that in BaK122(OPT). This means that the large SC gap is comparable in these compounds. On the other hand, the $1/T_1$ for Al-42622 decreases more steeply than in BaK122(OPT) as temperature falls  well below $T_{\rm c}$. This is primarily because the fraction of the density of states (DOS) at the Fermi level for FSs with a small SC gap, $r_{S}\equiv N_s^S/(N_s^L+N_s^S)$ is smaller for Al-42622 than for BaK122(OPT). 
Here $N_L$ and $N_S$ represent the respective DOSs with large and small SC gaps.
Actually, the result was well reproduced, assuming that $r_{S}\sim$0.1 for Al-42622 is smaller than $r_{S}\sim$0.3 for BaK122(OPT)\cite{Yashima}. It is deduced that $N_s^S$ is significantly smaller in Al-42622 than in BaK122(OPT). 
Note that in the simulation, a gap ratio $\Delta_0^S/\Delta _0^L$=0.35, a smearing factor $\eta $=0.14$\Delta_0^L$, and a coefficient of coherence factor $\alpha_{\rm c}\sim 0$ (see Ref.\cite{Nitta}) were used in BaK122(OPT)\cite{Yashima} for simplicity. Even when $N_s^S$=0 or $r_{S}$=0 is assumed,  the experiment can be also reproduced with $\eta\sim$0.3$\Delta _0^L$ larger than $\eta $=0.14$\Delta_0^L$ for $r_{S}\sim$0.1, as shown by the broken line in Fig.~\ref{simulation}(a). 

Next, we present an attempt to simulate a relaxation behavior below $T_{\rm c}$ for various Fe-based superconductors by changing the coefficient of 
coherence factor $\alpha_{\rm c}$. In this simulation, $\alpha_{\rm c}=1$ is assumed for sign-conserving {\it intraband} scattering and $\alpha_{\rm 
c}=-1$ for sign-nonconserving {\it interband} scattering. 
The value varies in the range $-1\le \alpha_{\rm c}\le 1$ dependent on the weight of their contribution in the nuclear relaxation process. 
In the previous studies on the heavily-overdoped LaFeAsO$_{1-x}$F$_{x}$(La1111(HOVD)) with $T_{\rm c}$=5 K\cite{Nitta} and optimally-doped La1111
(OPT) with $T_{\rm c}$=28 K\cite{MukudaNQR}, the experiments were reproduced with $\alpha _{\rm c}\sim 0.33$ for La1111(HOVD) and $\alpha _{\rm c}
\sim 0$ for La1111(OPT), as shown in Fig.~\ref{simulation}(b), which is attributed to the fact that the nesting condition of FSs becomes 
significantly worse in heavily overdoped regime. 
On the other hands, in Al-42622, AFM spin fluctuations develop significantly due to more dominant interband scattering than in the others. Relevant to this event, the 
experiment can be also reproduced by assuming {\it a negative value} of $\alpha_{\rm c}\sim -0.86$, as indicated in Fig.~\ref{simulation}(b), which 
contrasts with the previous studies.
Here, $r_{S}\le$0.1 and 2$\Delta/k_BT_{\rm c}$=6.1 were used along with other parameters used in La1111(OPT) with $T_{\rm c}$=28K\cite
{Yashima,Nitta}. It should be noted that the overall $T$ dependence of $1/T_1$ below $T_{\rm c}$ in Fe-based superconductors is consistently 
accounted for by the $s_\pm$-wave model with isotropic multiple gaps mainly through changing the coefficient of coherence factor $\alpha_{\rm c}$.  
We highlight the fact that the dominant {\it interband} scattering due to the nesting of hole and electron FSs is responsible for the marked enhancement 
of 2D AFM spin fluctuations and the sign-nonconserving {\it interband} scattering is responsible for the $T^7$-like reduction behavior in $1/T_1$ without the coherence peak below $T_{\rm c}$.

We have shown two possible simulations to reproduce the characteristic $T$ dependence of $1/T_1$ in the SC state of Al-42622. It is notable that, in either case, the assumption of $r_{S}\le$0.1 and $\alpha _{\rm c}\le 0$ were necessary, implying that the SC gaps on hole and electron FSs are nodeless with comparable sizes and the opposite signs. 
This event may be related to the band-calculation result that one of the hole FSs around $\Gamma$(0,0) disappears\cite{Miyake}. Nevertheless, AFM spin fluctuations are more significant in Al-42622 with $T_{\rm c}$=27~K than in La1111(OPT) with $T_{\rm c}$=28~K, but $T_{\rm c}$ is comparable for both. This result reveals that AFM spin fluctuations are not a unique factor for enhancing $T_{\rm c}$. Theoretically, within a spin-fluctuation mediated pairing theory on a five-orbital model, Usui {\it et al.} have claimed that the reduction of multiplicity of FSs in Al-42622 is a main reason why the $T_{\rm c}$ of Al-42622 is not so high even when AFM spin fluctuations are more remarkable than in existing Fe-based superconductors\cite{Usui}.  Since the FS topology is tuned by varying structural parameters such as pnictgen height and As-Fe-As bond angle $\alpha$, further systematic experiments are desired on a same series of Fe-based compounds. 

Finally, we comment on an $s_{++}$-wave model within orbital-fluctuation mediated pairing theory\cite{Kontani}. 
In general, the suppression of the coherence peak takes place in the strong-coupling regime of $s$-wave SC with relatively high $T_{\rm c}$ since strong-coupling effect causes $T_{\rm c}$ not only to increase, but also causes the lifetime of quasiparticles to shorten due to some damping effect\cite{Ohsugi,Kotegawa_MgB2}. 
For example, in a strong-coupling $s$-wave superconductor TlMo$_6$Se$_{7.5}$ with $T_c$=12.2 K, the coherence peak is suppressed  due to the phonon damping effect more significantly than in a weak-coupling one Sn$_{1.1}$Mo$_6$Se$_{7.5}$ with $T_c$=4.2 K\cite{Ohsugi}. A similar behavior was also observed for MgB$_2$ ($T_c\sim$40 K) and NbB$_2$ ($T_c$=5 K)\cite{Kotegawa_MgB2}. 
In Fe-based superconductors, the marked decrease of $1/T_1$ just below $T_{\rm c}$ is most significant in Al-42622 with $T_{\rm c}$=27 K among existing Fe-based superconductors, despite the fact that $T_{\rm c}$ is not so high relatively. It seems unlikely that the suppression of the coherence peak observed universally in most Fe-based superconductors can be systematically accounted for in terms of an $s_{++}$-wave model. Moreover, a non-magnetic impurity effect in La1111 compounds is not compatible with the $s_{++}$-wave state at all; While the crystal structure and electronic state are not modified by non-magnetic Zn substitution, the SC with $T_c=24$ K disappears by 3\% Zn substitution \cite{Guo,Kitagawa}.  

In conclusion, the $^{75}$As-NQR studies on (Ca$_4$Al$_2$O$_{6-y}$)(Fe$_2$As$_2$) with $T_{\rm c}$~=27~K have unraveled the development of 2D AFM spin fluctuations and pointed to the unconventional nodeless SC; The dominant interband scattering due to the nesting of hole and electron FSs is responsible for the marked enhancement of 2D AFM spin fluctuations and the sign-nonconserving interband scattering is responsible for the $T^7$-like reduction behavior in $1/T_1$ without the coherence peak below $T_{\rm c}$.  The $T$ evolution in $1/T_1$ in the SC state was consistently accounted for by the $s_{\pm }$-wave multiple gaps model. The present result also suggests that the DOS with a small SC gap is totally reduced in association with the disappearance of some part of Fermi surfaces.  From the fact that $T_{\rm c}=27$~K in this compound is comparable to $T_c$=28~K in the optimally-doped LaFeAsO$_{1-y}$ in which AFM spin fluctuations are not dominant, we remark that AFM spin fluctuations are not a unique factor for enhancing $T_c$, but a condition for optimizing SC should be addressed from the lattice structure point of view.  


We thank K. Kuroki for valuable comments. This work was supported by a Grant-in-Aid for Specially Promoted Research (20001004) and by the Global COE Program (Core Research and Engineering of Advanced Materials-Interdisciplinary Education Center for Materials Science) from the Ministry of Education, Culture, Sports, Science and Technology (MEXT), Japan.


\end{document}